\documentclass{aa}  
\usepackage{graphicx}
\usepackage{natbib}
\usepackage{hyperref}
\titlerunning{}

\usepackage{color}

\newcommand{\kms}{\mbox{$\mbox{km\,s}^{-1}$}\,}

\begin{document}

\title{Calibrating the surface brightness - color relation for late-type red giants
stars in the visible domain using VEGA/CHARA interferometric observations}

\titlerunning{Calibrating the surface brightness - color relation for late-type stars}
\authorrunning{Nardetto et al. }
\author{N. Nardetto \inst{1} 
\and A. Salsi\inst{1}
\and D. Mourard \inst{1} 
\and V. Hocd\'e\inst{1}
\and K. Perraut \inst{2} 
\and A. Gallenne\inst{1,3,4,5}  
\and A. M\'erand \inst{6} 
\and D. Graczyk \inst{3, 4} 
\and G. Pietrzynski \inst{4}
\and W. Gieren\inst{3} 
\and P. Kervella \inst{7}    
\and R. Ligi \inst{8} 
\and A. Meilland \inst{1} 
\and F. Morand \inst{1} 
\and P. Stee \inst{1} 
\and I. Tallon-Bosc \inst{9} 
\and T. ten~Brummelaar\inst{10,11} 
}
\institute{Universit\'e C\^ote d'Azur, Observatoire de la C\^ote d'Azur, CNRS, Laboratoire Lagrange, France
\and Univ. Grenoble Alpes, CNRS, IPAG, 38000 Grenoble, France
\and Departamento de Astronom\'ia, Universidad de Concepci\'on, Casilla 160-C, Concepci\'on, Chile 
\and Nicolaus Copernicus Astronomical Center, Polish Academy of Sciences, ul. Bartycka 18, PL-00-716 Warszawa, Poland. 
\and Unidad Mixta Internacional Franco-Chilena de Astronom\'ia (CNRS UMI 3386), Departamento de Astronom\'ia, Universidad de Chile, Camino El Observatorio 1515, Las Condes, Santiago, Chile
\and European Southern Observatory, Alonso de C\'ordova 3107, Casilla 19001, Santiago 19, Chile 
\and LESIA (UMR 8109), Observatoire de Paris, PSL, CNRS, UPMC, Univ. Paris-Diderot, 5 place Jules Janssen, 92195 Meudon, France 
\and Osservatorio Astronomico di Brera, Via E. Bianchi 46, 23807 Merate, Italy.
 \and Universit\'e de Lyon, Universit\'e Lyon 1, Ecole Normale Sup\'erieure de Lyon, CNRS, Centre de Recherche Astrophysique de Lyon UMR5574, F-69230, Saint-Genis-Laval, France 
 \and Georgia State University, P.O. Box 3969, Atlanta GA 30302-3969, USA  \and CHARA Array, Mount Wilson Observatory, 91023 Mount Wilson CA, USA }

\date{Received ... ; accepted ...}

\abstract{The surface brightness - color relationship (SBCR) is a poweful tool for determining the angular diameter of stars from photometry. It was for instance used to derive the distance of eclipsing binaries in the Large Magellanic Cloud (LMC), which led to its distance determination with an accuracy of 1\%.
} {We calibrate the SBCR for red giant stars in the  $2.1 \leq V-K \leq 2.5 $ color range using homogeneous VEGA/CHARA interferometric data  secured in the visible domain, and compare it to the relation based on infrared interferometric observations, which were used to derive the distance to the LMC.} {Observations of eight G-K giants were obtained with the VEGA/CHARA instrument. The derived limb-darkened angular diameters were combined with a homogeneous set of infrared magnitudes in order to constrain the SBCR.} {The average precision we obtain on the limb-darkened angular diameters of the eight stars in our sample is 2.4\%. For the four stars in common observed by both VEGA/CHARA and PIONIER/VLTI, we find a  $1 \sigma$ agreement for the angular diameters. The SBCR we obtain in the visible has a dispersion of 0.04 magnitude and is consistent with the one derived in the infrared (0.018 magnitude).}{The consistency of the infrared and visible angular diameters and SBCR reinforces the result of 1\% precision and accuracy recently achieved on the distance of the LMC using the eclipsing-binary technique. It also indicates that it is possible to combine interferometric observations at different wavelengths when the SBCR is calibrated.}

\maketitle

\section{Introduction}\label{s_Introduction}
\vspace{1cm}

In the era of precision cosmology, it is essential to determine the Hubble constant to an accuracy of 2\% or better \citep{komatsu11, freedman10}. Recently, \citet{riess19} achieved a 1.9\% precision by relying on the distance to the Large Magellanic Cloud (LMC), which is the best anchor point for the cosmic distance scale \citep{schaefer08, walker12, riess16}. Late-type eclipsing-binary systems provide the opportunity of measuring accurate distances by combining the linear diameter (derived from the light curve and velocimetry) and angular diameters (derived from a surface-brightness color relation, SBCR) of their components.

Based on the observation of some 35 million stars in the LMC for more than 20 years during the Optical Gravitational Lensing Experiment (OGLE) \citep{udalski08}, a few dozen extremely scarce long-period eclipsing systems composed of late-type clump giants were cataloged \citep{graczyk11, pawlak16}. 
In the course of the Araucaria project \citep{gieren05}, \citet{gp13} studied 8 systems and derived a distance to the LMC at the 2.2\% level. This distance relied on the SBCR derived by \citet{dibenedetto05} and thus was completely dominated by the error of 2\% (or 0.03 magnitude) on this relation. Then, \citet{laney12} secured uniform and precise K-band photometry (precision of about $\simeq$0.01 mag) for a large sample of nearby very bright red giants located in a very well defined and quiet evolutionary phase of the core-helium burning (red clump). In addition, \citet{gallenne18} secured observations of 48 of these red-clump stars in the H band with the Precision Integrated Optics Near-infrared Imaging ExpeRiment (PIONIER, \citet{lebouquin11}) installed at the Very Large Telescope Interferometer. These measurements allowed deriving an SBCR with an rms of 0.018 magnitude. This relation was used to derive the individual angular diameters of the 20 eclipsing binaries at the 0.8\% precision level, which finally led to a distance of the LMC that is accurate to 1\% \citep{gp19}.

The purpose of this paper is to calibrate the SBCR in the visible domain in order to compare it to the one derived in the H band. For this, we observed a subsample of eight stars in the visible domain with interferometry, using the Visible spEctroGraph and polArimeter (VEGA) beam combiner \citep{mourard09, mourard11} operating at the focus of The Center for High Angular Resolution Astronomy (CHARA) array \citep{ten05}, which is located at the Mount Wilson Observatory. In Sect.~\ref{data} we describe the VEGA/CHARA interferometric observations and provide the limb-darkened angular diameters of the eight stars in our sample. Sect.~\ref{sb} is dedicated to the calibration of the SBCR. The results are then discussed in Section~\ref{sect_discussion}, and we conclude in Section~\ref{sect_conclusion}.  

\section{VEGA/CHARA observations of eight late-type stars}\label{data}
We selected eight late-type stars with existing infrared photometry by \citet{laney12}, with a $(V-K)_\mathrm{0}$ color index ranging from 2.12 to 2.43, which corresponds to the index that was used to derive the distance to the LMC (2.05-2.75). They are red giants stars ($\delta$ between -9$^{o}$ to 12$^{o}$) with spectral types ranging from G8 to K2. They have a visual magnitude $m_\mathrm{V}$ ranging from 5.8 to 6.2, which is well below the limiting magnitude of VEGA. They are also bright in the K band (with  $m_\mathrm{K} < 3.9$), which allows us to track the group delay simultaneously with the infrared CLIMB combiner \citep{sturmann10}.  We observed our sample from 2013 July   to 2014 October using different suitable telescopes available on the CHARA array. A summary of the observations is given in Table~\ref{Tab2}.

\begin{table}
\begin{center}
\caption[]{Summary of the observing log. All the details are given in Tab.~\ref{Tabapx.1}. $N$ corresponds to the number of visibility measurements for each star. The reference stars are also indicated (cf. Tab. \ref{Tab3}).}
\label{Tab2}
\begin{tabular}{p{1.5cm}p{3.3cm}p{0.cm}p{2.8cm}c}
\hline\hline
 star        &   telescope configurations & $N$  & reference stars \\
\hline

HD11037  &  E1E2W2, W1W2S2 &  4    & C1, C2      \\ 
HD13468  &  E1E2W2, W1W2E2   &  7    & C3     \\ 
HD22798  &  E1E2W2, W1W2S2   &  4    & C4       \\
HD23526  &  E1E2W2, W1W2S2  &  9   & C4       \\
HD360   & S1S2W2 & 7 & C5, C6  \\
HD40020  &  E1E2, W1W2   &  6    & C7, C8       \\
HD43023  &  E1E2W2, W1W2S2    &  7   & C9, C10, C11, C12      \\ 
HD5268  & W2W1S2      &   2       &  C5, C6   \\

 \hline
\end{tabular}
\end{center}
\end{table}

\begin{table}
\begin{center}
\caption{Reference stars and their parameters, including the spectral type, the visual magnitude (m$_\mathrm{V}$), and the predicted uniform-disk angular diameter (in mas) together with its corresponding uncertainty derived from the JMMC {\it SearchCal} software \citep{bonneau06, lafrasse10}. The uncertainties on the reference stars are about 7\%, which is conservative. }\label{Tab3}
\begin{tabular}{p{0.6cm}p{1.4cm}p{1.4cm}p{0.6cm}c}
\hline\hline
 & reference stars & spectral type      & m$_\mathrm{V}$  & $\theta_\mathrm{UD}[R]$  \\
      &   &   & [mag]  &  [mas] \\
\hline
C1  & HD18604     &  B6III    &  $4.70$    &  $ 0.259 \pm 0.018$        \\
C2  & HD224926     &   B7III-IV   &   $5.10$    &   $0.192 \pm 0.014$   \\
C3  & HD15633     &   A3V  &  $6.01$     &   $0.252 \pm 0.018$  \\ 
C4  & HD23363    &  B7IV    &     $5.25$  &   $0.201 \pm 0.014$ \\
C5  & HD219402    &  A2V    &    $5.55$   &   $0.249 \pm 0.018$ \\ 
C6  & HD6530   &  A0V    &    $ 5.58$   &  $0.234 \pm 0.017$    \\ 
C7  & HD34203   &       A0V    & $ 5.52$      &  $ 0.221 \pm 0.016$     \\
C8  & HD30034   &  F0V    &  $ 5.38$     &  $0.407 \pm 0.029$   \\
C9  & HD43445   &   B9V   &   $5.00$    &   $0.239 \pm 0.017$   \\
C10 & HD32249   &  B3IV    &    $ 4.81$   &   $0.176 \pm 0.012$ \\ 
C11  & HD46487   &  B5IV/V    &   $ 5.08$    &   $0.180 \pm 0.013$      \\ 
C12  & HD34863   &  B7/8V    &  $5.28$     &   $0.181 \pm 0.013$        \\
\hline

\end{tabular}
\end{center}
\end{table}

For these observations, we used the medium spectral resolution mode of VEGA ($R \simeq 5000$) and the standard VEGA pipeline in order to calibrate the squared visibilities in spectral bands of 15 or 20 nm \citep{mourard09, mourard11, ligi13}.  The integration time of our observations is of 500 seconds. For each calibrated visibility, the statistical and systematic calibration errors are given separately in Tab.~\ref{Tabapx.1}. The systematic uncertainty stems from the uncertainty on the calibrator diameter, which is given in Tab.~\ref{Tab3}.  To perform the model fitting, we used a JMMC\footnote{\url{http://www.jmmc.fr/}} tool, {\it LITpro} \citep{tallonbosc08}. The systematic and statistical errors are considered separately in the fitting procedure and are propagated until the final error on the surface brightness.
In Fig.~\ref{fig.V2} and \ref{fig.V2b},  the calibrated visibilities are plotted as a function of the spatial frequency together with the corresponding (u,v) coverage for each star. We  fit these calibrated visibilities by a uniform disk, where the so-called uniform-disk angular diameter ($\theta_\mathrm{UD}$) is the only parameter. The results are given in Tab.~\ref{tab.param}.  We then converted these $\theta_\mathrm{UD}$ diameters into limb-darkened ($\theta_\mathrm{LD}$) angular diameters. A common and convenient approach is to use a linear law for the continuum-intensity profile of the star,  defined by $I(\cos(\alpha))=1-u+u\cos(\alpha)$, where $u$ is the linear limb-darkening coefficient of the star in the interferometric wavelength band of observation \citep{claret11}. $\alpha$ is the angle between the normal of the star and the line of sight. The limb-darkened angular diameter is then calculated using $\theta_{\mathrm{LD}}=\theta_{\mathrm{UD}}\left[\frac{(1-\frac{u}{3})}{(1-\frac{7u}{15})}\right]^{\frac{1}{2}} $ \citep{brown74}.  In the Claret tables, $u$ is given as a function of the effective temperature $T_\mathrm{eff}$, the surface gravity $\log g$, the metallicity $Z$ and the microturbulence velocity $V_\mathrm{t}$,   in several photometric bands, including the R band ($\lambda_\mathrm{eff}=670$nm) and the I band ($\lambda_\mathrm{eff}=856$nm). The fundamental parameters of the stars are given in Tab.~\ref{tab.uR} together with the rounded values used as an input in the Claret tables. After they are extracted from these tables, the limb-darkened coefficients in the R and I bands were interpolated at the typical wavelength band of our observations, that is, 710 mn, in order to derive the corresponding linear limb-darkening coefficient ($u_\mathrm{[710]}$). The resulting $u_{\mathrm [710]}$ parameters are given in Tab.~\ref{tab.param} together with the corresponding limb-darkened angular diameters. The reduced $\chi^2$ ranges from 0.4 to 2.9. 

Changing the effective temperature by 500 K and/or the surface gravity by 0.5 have an effect on the limb-darkening angular diameter of 1.3\%.  The grid in effective temperature has a step of 250 K, thus the largest error we can make on the temperature is 125 K, which corresponds to an additional uncertainty on the limb-darkened diameter of about 0.3\%. However, the fact that the derived diameters are weakly sensitive to the input parameters of the limb-darkening law does not exclude that they might be sensitive to the limb-darkening law itself or to the method. We also tested the method by fitting the limb-darkened angular diameter with the {\it limb\_linear} function of the {\it LITpro} tool, that is, by fixing the limb-darkening coefficient to its value indicated in Tab.~\ref{tab.uR}. The derived limb-darkened angular diameters are 0.4 to 0.8\% larger (except for HD 23526, for which we find 1.2\%) than those indicated in Tab.~\ref{tab.uR} based on the approximate analytic conversion law. This bias is larger than what has been stated by  \citet{brown74}, which is an approximation of $<0.4$\% if $u_{\mathrm R}$ ranges from 0.5 to 1.0. 
The uncertainties are strictly the same, however. For the limb-darkening law itself, we also used the dedicated SATLAS models by
\citet{neilson13a, neilson13b} to derive the limb-darkened diameters using the same input physical parameters as those indicated in Tab.~\ref{tab.uR}. With this approach, we find diameters that are about 0.4\% larger than the values presented in Tab.~\ref{tab.param}. These tests are interesting and should be considered with caution when a limb-darkened angular diameter is derived at the sub-percent level, but our statistical precision here is about 2.4\% and the common analytical approach leads to a bias on our SBCR calibration at a far lower level than our statistical precision.

\begin{figure}[htbp]
\begin{center}
\resizebox{1.0\hsize}{!}{\includegraphics[clip=true]{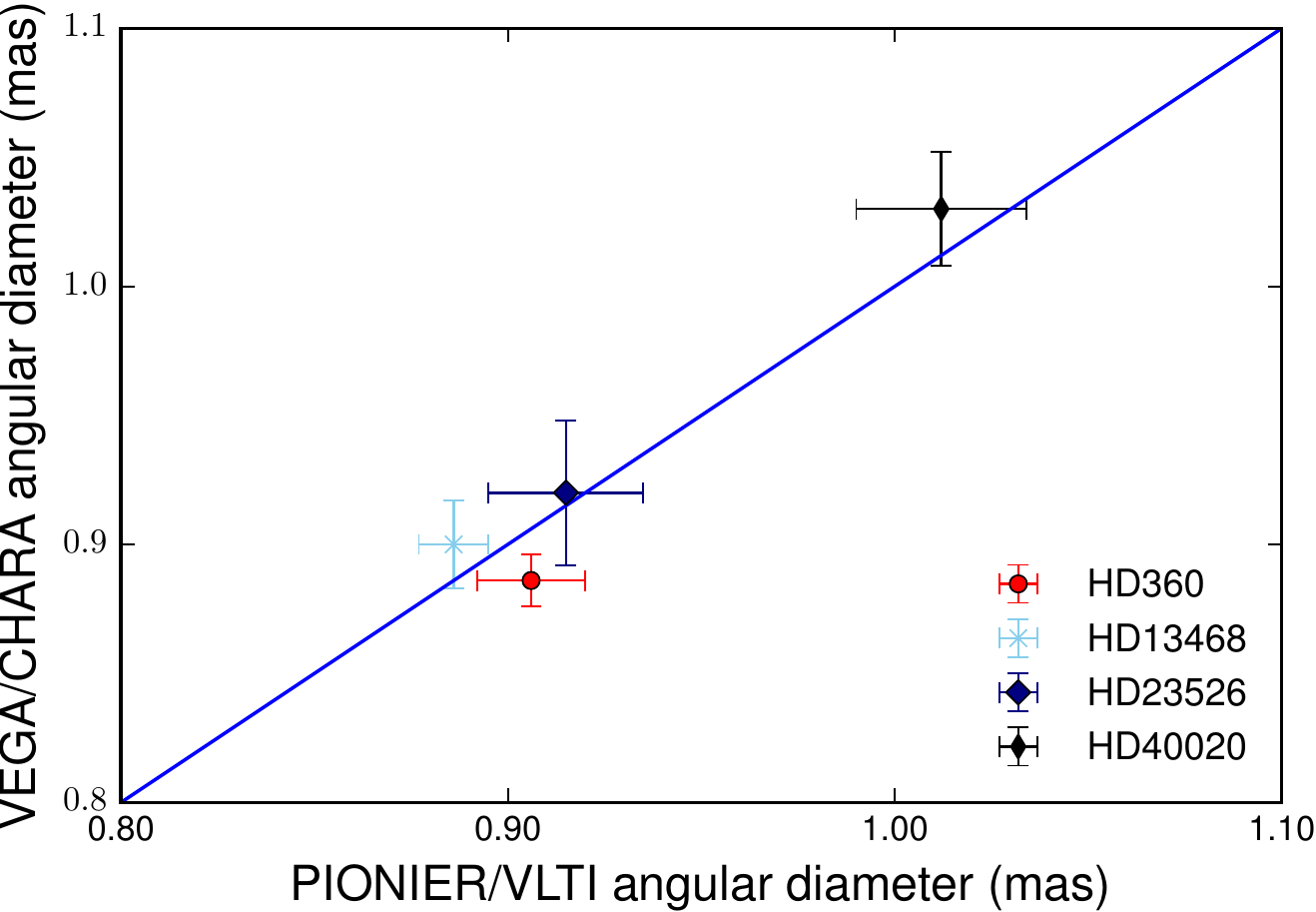}}
\end{center}
\vspace{-0.5cm}
\caption{Angular diameters derived from VEGA/CHARA compared to those from PIONIER/VLTI. The blue line marks the 1:1 relation.} \label{Fig.diam}
\end{figure}

Our sample includes four stars that are in common with samples of \citet{gallenne18} and \citet{gp19}: HD13468, HD23526, HD360, and HD40020.  Their PIONIER/VLTI angular diameters are listed in Tab.~\ref{tab.param}. This is an interesting opportunity to compare the angular diameters derived from CHARA/VEGA and VLTI/PIONIER (see Fig.~\ref{Fig.diam}). We find that the CHARA/VEGA angular diameters are consistent within their uncertainties with those from VLTI/PIONIER. Visible (this work) and infrared \citep{gallenne18} interferometric observations are thus in agreement, which means that the conversions from UD into LD angular diameters are at least consistent in the two wavelength domains  ($R$ and $H$) at the sub-percent level. Moreover, it indicates that there is negligible systematics on the instrumental point of view in the determination of the angular diameters. Both instruments might be biased in exactly the same way (for the four stars), but this is highly unlikely. The calibrators used in the VLTI/PIONIER analysis are giants with spectral types ranging from G8 to K2, which is significantly colder and more resolved than the calibrators we used in our analysis in the visible domain (Tab.~\ref{Tab3}). We therefore expect that the systematics on the squared visibilities (Tab.~\ref{Tab3}) that stem from the uncertainties on the calibrators are significantly lower in the visible than in the infrared.

\begin{table*}
\begin{center}
\caption[]{Angular diameters obtained with VEGA/CHARA and the corresponding surface brightnesses. For each star, we consider an uncertainty of 0.01 magnitude for $m_\mathrm{V}$ and $m_\mathrm{K}$ \citep{laney12}. The angular diameter of stars marked with an asterisk is based on PIONIER/VLTI observations provided in the extended data Table 2 of \citet{gp19}. This is  reported here in the last column.  }
\label{tab.param}
\setlength\tabcolsep{3pt}
\begin{tabular}{lccccccccc|c}
\hline\hline 
Star &  $m_\mathrm{V}$ & $m_\mathrm{K}$ &  $A_\mathrm{V}$ & $(V-K)_0$ & $\theta_{\mathrm{UD}}    {}^{\pm \sigma_\mathrm{stat}}_{\pm \sigma_\mathrm{syst}} $  & $\chi_\mathrm{red}^{2}$ & $u_\mathrm{[710]}$ & $\theta_{\mathrm{LD}} {}^{\pm \sigma_\mathrm{stat}}_{\pm \sigma_\mathrm{syst}}$ & $S_{\mathrm{v}} {}^{\pm \sigma_\mathrm{stat}}_{\pm \sigma_\mathrm{syst}}$ & $\theta_{\mathrm{LD}}{}^{\pm \sigma_\mathrm{stat}}_{\mathrm{PIONIER}}$ \\
  &  [mag] &[mag] & [mag]&  [mag]& [mas] &  &  & [mas] & [mag] & [mas] \\
\hline
HD11037 &  5.910$_{\pm 0.010}$  & 3.666$_{\pm 0.010}$ & 0.031   & $2.216$ &$0.841^{\pm0.018}_{\pm 0.002}$ & $1.3$ &$0.631$  &$0.890^{\pm0.019}_{\pm 0.002}$ & $5.626^{\pm0.047}_{\pm 0.011}$ &  \\ 
\smallskip
HD13468$^{*}$  & 5.940$_{\pm 0.010}$  & 3.666$_{\pm 0.010}$ &  0.028  & $2.248$ & $0.851^{\pm0.016}_{\pm 0.002}$ & $1.6$ &  $0.628$  &$0.900^{\pm0.017}_{\pm 0.002}$ & $5.683^{\pm0.042}_{\pm 0.011}$ & $0.886^{\pm0.009}$  \\ 
\smallskip
HD22798 &  6.220$_{\pm 0.010}$  & 3.880$_{\pm 0.010}$ &  0.012  &  $2.329$&$0.749^{\pm0.020}_{\pm 0.002}$ & $2.9$ & $0.631$ &$0.792^{\pm0.021}_{\pm 0.002}$ & $5.701^{\pm0.058}_{\pm 0.011}$ & \\
\smallskip
HD23526$^{*}$ &  5.910$_{\pm 0.010}$    & 3.634$_{\pm 0.010}$ & 0.053   & $2.228$ &$0.869^{\pm0.025}_{\pm 0.001}$ & $0.6$ & $0.631$ &$0.920^{\pm0.027}_{\pm 0.001}$ & $5.676^{\pm0.064}_{\pm 0.010}$ & $0.915^{\pm0.020}$ \\ 
\smallskip
HD360$^{*}$   &    5.990$_{\pm 0.010}$   & 3.653$_{\pm 0.010}$&   0.028   &  $2.311$&$0.835^{\pm0.009}_{\pm 0.002}$ & $0.5$ & $0.658$ &$0.886^{\pm0.010}_{\pm 0.002}$ & $5.699^{\pm0.026}_{\pm 0.011}$ & $0.906^{\pm0.014}$ \\
\smallskip
HD40020$^{*}$  &   5.890$_{\pm 0.010}$  &3.419$_{\pm 0.010}$ &   0.040  &  $2.434$&$0.969^{\pm0.024}_{\pm 0.005}$ & $0.4$ & $0.668$ & $1.030^{\pm0.025}_{\pm 0.005}$  & $5.913^{\pm0.053}_{\pm 0.014}$ & $1.012^{\pm0.022}$ \\ 
HD43023  & 5.830$_{\pm 0.010}$    &3.704$_{\pm 0.010}$ & 0.003    &  $2.123$ &$0.796^{\pm0.012}_{\pm 0.001}$ & $1.1$ & $0.631$ &$0.842^{\pm0.014}_{\pm 0.001}$ & $5.453^{\pm0.037}_{\pm 0.010}$ &  \\
\smallskip
HD5268  &  6.150$_{\pm 0.010}$  &3.910$_{\pm 0.010}$ &   0.043   &  $2.200$ &$0.725^{\pm0.033}_{\pm 0.005}$ & $0.5$ & $0.628$ &$0.767^{\pm0.035}_{\pm 0.005}$  & $5.530^{\pm0.099}_{\pm 0.017}$ & \\
\hline
\end{tabular}
\end{center}
\end{table*}

\begin{table*}
\begin{center}
\caption[]{ Fundamental parameters of the stars in our sample based on spectroscopy and combined with photometry for stars in \citet{luck07}.}
\label{tab.uR}
\begin{tabular}{lcccccccccl}
\hline \hline
Star &  Sp. Type & $T_\mathrm{eff}$ & $\log g$ & Z & $V_t$ & round  & round  & round  & round  & references\\
&   & &  & &  &$T_\mathrm{eff}$ &  $\log g$ &  Z &  $V_t$ & \\
            &      &    [K]   &   &    & [\kms] &   [K]  & &  & [\kms] \\
\hline
HD11037 & K0III & 4976  &       2.85    &       -0.10   &       1.5     &       5000    &       3.0     &       0.0     &       1.0     & \citet{luck07} \\ 
HD13468 & K0III  & 4940 &       2.59    &       -0.17   &       1.0     &       5000    &       2.5     &       0.0     &       1.0     & \citet{jones11} \\ 
HD22798 &       K0III   & 4905  &       2.99    &       0.24    &       1.0     &       5000    &       3.0     &       0.2     &       1.0     & \citet{soubiran08} \\ 
HD23526 &       G9III & 4935    &       2.81    &       -0.12   &       1.0     &       5000    &       3.0     &       0.0     &       1.0     &  \citet{luck07} \\
HD360   &       G9III & 4741    &       2.73    &       -0.05   &       1.3     &       4750    &       3.0     &       0.0     &       1.0     & \citet{liu07}  \\ 
HD40020 &       K2III & 4752    &       2.67    &       0.17    &       2.0     &       4750    &       2.5     &       0.2     &       2.0     & \citet{luck07}  \\
HD43023 &       K0III & 5105    &       3.08    &       -0.06   &       1.5     &       5000    &       3.0     &       0.0     &       1.0     & \citet{luck07}  \\ 
HD5268  &       G8III  & 4904   &       2.35    &       -0.57   &       1.0     &       5000    &       2.5     &       0.0     &       1.0     &  \citet{jones11} \\ 
\hline
\end{tabular}
\end{center}
\end{table*}

\begin{figure*}[htbp]
\begin{center}
\resizebox{0.65\hsize}{!}{\includegraphics[clip=true]{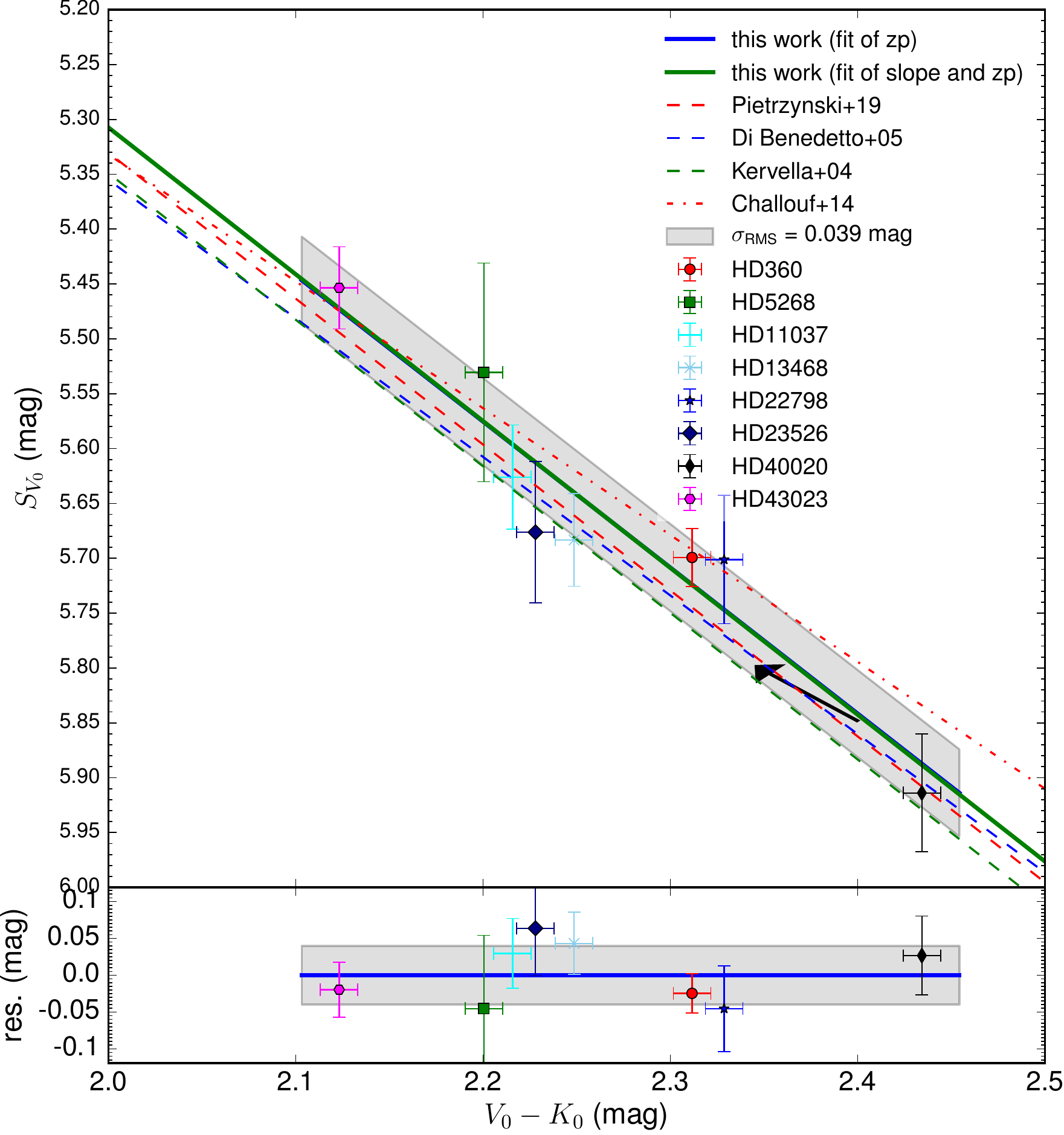}}
\end{center}
\vspace{-0.5cm}
\caption{Surface brightness -- color relation derived from VEGA/CHARA data of eight late-type stars with a homogenous set of infrared photometry \citep{laney12}. The black arrow shows the effect of an additional extinction of 0.1 magnitude on a measurement. The gray zone corresponds to the rms.} \label{Fig.Sv}
\end{figure*}


\section{Calibrating the surface brightness relation}\label{sb}

The surface brightness of a star is linked to its visual intrinsic dereddened magnitude $m_\mathrm{V_{\mathrm{0}}}$ and its limb-darkened angular diameter $\theta_{\rm LD}$ by the following relation: $S_{\mathrm{V}}=m_\mathrm{V_{\mathrm{0}}}+5\log \theta_{LD}$. 
In order to derive $m_\mathrm{V_{\mathrm{0}}}$, we first selected the apparent $m_\mathrm{V}$ magnitude from the main Hipparcos and Tycho catalog \citep{Hip97} expressed in the Johnson system \citep{johnson66}. For the dereddening of these magnitudes, we used $ m_\mathrm{V_{\mathrm{0}}}=m_\mathrm{V}-A_{V}$ , where $A_{V}$ is either taken directly from \citet{gp19} for the four targets we have in common or based on Stilism\footnote{\url{https://stilism.obspm.fr/}} \citep{lallement14,capitanio17}. \citet{gp19} used the 2D reddening maps of \citet{schlegel98} and \citet{schlafly11}. These maps give a total extinction in a given line of sight, while
the reddening to a given star is only a small fraction of it. To calculate this fraction, they used a simple exponential model of dust distribution in the Milky Way, the Hipparcos parallaxes, and the local extinction-free bubble
radius of 40 pc \citep{Suchomska15}.  Stilism is a 3D maps of the Galactic interstellar medium based on a combination of various methods and databases.  The extinction we used for each star in our sample is indicated in Tab.~\ref{tab.param}, and the effect of these extinctions on the SBCR is discussed in the next section. Then the color follows with $(V-K)_{0}=(V-K)-A_V-A_K$ , where  $A_\mathrm{K}=0.089 A_\mathrm{V}$ \citep{nishiyama09}. The $m_\mathrm{K}$ magnitudes are homogeneously derived from \citet{laney12}. The $S_\mathrm{V}$ values are given in Table~\ref{tab.param}. 


In order to fit our SBCR, we used the same formalism as
\citet{gp19},

\begin{equation}\label{Eq_SBCR}
 S_{\mathrm{V}} =\alpha   [(V-K)_\mathrm{0} - 2.405] + \beta
,\end{equation}
where $\alpha$ and $\beta$ are the slope and zero-point of the relation, respectively. The value $2.405$ is the average color of the stars that were used to built the SBCR in \citet{gp19}. We kept this value in our fit to allow a direct comparison. We first fit both parameters and then assumed the slope from the SBCR of \citet{gp19}. The statistical uncertainty on the  magnitudes $m_\mathrm{V}$ and $m_\mathrm{K}$, on the extinction $A_\mathrm{V}$ , and on the surface brightness $S_{\mathrm{v}} $ were taking into account in the fitting process. Results are summarized in 
Tab.~\ref{tab.res}. We obtain very similar results when we fit one or two parameters (1.7\% difference on the slope and 0.05\% on the zero-point), as shown by Fig.~\ref{Fig.Sv}.
The SBCRs from \citet{kervella04}, \citet{dibenedetto05}, and \citet{challouf14} are shown for comparison.



\begin{table*}
\begin{center}
\caption[]{Fit of the SBCR. Slope and zero-point of the SBCR were first fit using Eq.~\ref{Eq_SBCR}. We then assumed the slope from the SBCR of \citet{gp19} (indicated by an asterisk). The reduced $\chi^2$ is 0.4 in both cases. 
 }
\label{tab.res}
\begin{tabular}{p{3.6cm}p{2.1cm}p{2.1cm}p{2.1cm}}
\hline\hline
Reference       &  $\beta$ & $\alpha$  & rms\\
\hline
Pietrzynski et al. (2019) & $5.869_\mathrm{\pm0.003}$  & $1.330_\mathrm{\pm0.017}$ & 0.018 \\
this work & $5.849_\mathrm{\pm0.027}$  &  $1.338_\mathrm{\pm0.160}$ & 0.039 \\
this work & $5.848_\mathrm{\pm0.013}$  &  $1.330^{\star}$ & 0.039 \\

\hline
\end{tabular}
\end{center}
\end{table*}

\section{Discussion}\label{sect_discussion}

Our relation has an rms of 0.04 magnitude (Tab.~\ref{tab.res}) and is consistent with all other SBCRs presented in Fig.~\ref{Fig.Sv}, except for that of \citet{challouf14}, in particular for a $(V-K)$ color larger than 2.3 magnitude. This is explained by the fact that the relation by \citet{challouf14} is nonlinear, was based on 132 stars of all classes, and was fit over a very wide range of color from early to late types. We note that this relation was dedicated to O, B, A stars. 

Conversely, the relation by \citet{gp19}, which has an rms of 0.018 magnitude, is consistent with all SBCRs (except for that of \citet{challouf14}), including the SBCR in this work. These relations are indeed consistent (within 1\%), even though different methods were used. The relation by \citet{gp19} is based on the observation of 48 giant stars in the H band with the PIONIER/VLTI \citep{gallenne18}, while the relation by \citet{kervella04} is based on a set of angular diameters of dwarfs and subgiant stars derived from different optical and infrared interferometers: 
NPOI, NII (Narrabri Intensity Interferometer, \citet{brown67}), Mark III \citep{shao88}, PTI (Palomar Testbed Interferometer, \cite{colavita99}), and VINCI \citep{kervella04g}. The relation by \citet{dibenedetto05} is based on the NPOI (Navy Prototype Optical Interferometer, \citet{armstrong98}) observations by \citet{nordgren01} of 24 giants and three dwarfs obtained in the optical domains (wavelength of reference of 740 nm). Thus, following Salsi et al. (2020, in preparation), who found a different SBCR depending on the spectral type and class of the stars, we can consistently compare the relations by \citet{gp19}, \citet{dibenedetto05} and ours, which are all based on giants. On the color range considered, the relation in this work (rms of 0.039 mag) is consistent with the relation by \citet{dibenedetto05} (rms of 0.04 mag) and with the relation by \citet{gp19} (rms of 0.018 mag). 

It is well known that the extinction has a negligible effect on the SBCR because the color and surface brightness are sensitive to it in almost the same way. The only difference comes from the extinction in K, which is about 10\% of the extinction in V.  Adding an extinction of 0.1 magnitude on a measurement, for instance, results in a brighter color of the star of about 0.054 magnitude, and similarly, in a brighter surface-brightness of 0.05 magnitude. In order to quantify the effect of extinction on the SBCR, we arbitrarily changed the absorption ($A_\mathrm{V}$) of all stars in our sample. We find that a higher extinction of 0.1 magnitude results in an increase in zero-point of the SBCR by 0.045 magnitude, which corresponds to the rms of our SBCR. This offset becomes 0.048 magnitude when we use the older relationship $A_\mathrm{K}=0.114 A_\mathrm{V}$ from \citet{cardelli89} instead of the relation we considered in Sect. 3  ($A_\mathrm{K}=0.089 A_\mathrm{V}$) from \citet{nishiyama09}. This means that even if the $A_\mathrm{V}$ values of the star in the sample were all under-estimated by 0.1 magnitude (which is highly unlikely), our SBCR would still be consistent with the SBCR by \citet{gp19}. 

\section{Conclusion}\label{sect_conclusion}

We observed eight red clump stars (late-type giants) with the VEGA/CHARA instrument and derived their angular diameters with an average precision of 2.4\%. By combining these diameters with a homogeneous set of infrared photometry \citep{laney12}, we derived the slope and zero-point of the SBCR with a residual dispersion of 0.04 magnitude (or 1.8\% precision in terms of angular diameter). The zero-points of the SBCR derived in the H band and in the visible domain are consistent at the 0.4\% level (or 0.5$\sigma$). The angular diameters of four stars in our sample are consistent with those found by VLTI/PIONIER,  which shows that the systematics of both instruments are under control
It also indicates that it is possible to combine  interferometric measurements in different bands when the SBCR is calibrated. All these results confirm the recent claim of a 1\% precision and accuracy on the LMC distance by \citet{gp19}. 

\begin{acknowledgements}
We acknowledge M. Schultheis for fruitful discussions about the extinction of stars. This research has made use of the SIMBAD and VIZIER\footnote{Available at http://cdsweb.u- strasbg.fr/} databases at CDS, Strasbourg (France) and of electronic bibliography maintained by the NASA/ADS system. This research has made use of the Jean-Marie Mariotti Center \texttt{LITpro} service co-developped by CRAL, IPAG and LAGRANGE\footnote{LITpro software available at http://www.jmmc.fr/litpro}. This research has made use of the Jean-Marie Mariotti Center \texttt{Aspro} service \footnote{Available at http://www.jmmc.fr/aspro}.  This work was supported by the "Programme National de Physique Stellaire (PNPS) of CNRS/INSU co-funded by CEA and CNES. This work was supported by the "Action Sp\'ecifique pour la Haute R\'esolution Angulaire (ASHRA) of CNRS/INSU co-funded by CNES. The CHARA Array is funded by the National Science Foundation through NSF grants AST-0606958 and AST-0908253 and by Georgia State University through the College of Arts and Sciences, as well as the W. M. Keck Foundation. R.L. has received funding from the European Union's Horizon 2020 research and innovation programme under the Marie Sklodowska-Curie grant agreement n. 664931.  W.G. and G.P. gratefully acknowledge financial support for this work from the BASAL Centro de Astrofisica y Tecnologias Afines (CATA) AFB-170002. We acknowledge financial support for this work from ECOS-CONICYT grant C13U01. The  authors  acknowledge  the  support  of  the  FrenchAgence  Nationale  de  la  Recherche  (ANR),  under  grant  ANR-15-CE31-0012-01 (project UnlockCepheids).  Support from the Polish National ScienceCentre grants MAESTRO UMO-2017/26/A/ST9/00446 and from the IdPII 2015 0002 64 grant of the Polish Ministry of Science and Higher Education is also acknowl-edged. The research leading to these results has received funding from the European Re-search Council (ERC) under the European Unions Horizon 2020 research and innovation program (grant agreement No 695099).  
\end{acknowledgements}
 
\bibliographystyle{aa}  
\bibliography{bibtex_nn} 

\begin{appendix}
\section{Plot of the CHARA/VEGA squared visibility measurements.}
\begin{figure*}
 \begin{center}
 \resizebox{0.35\hsize}{!}{\includegraphics[clip=true]{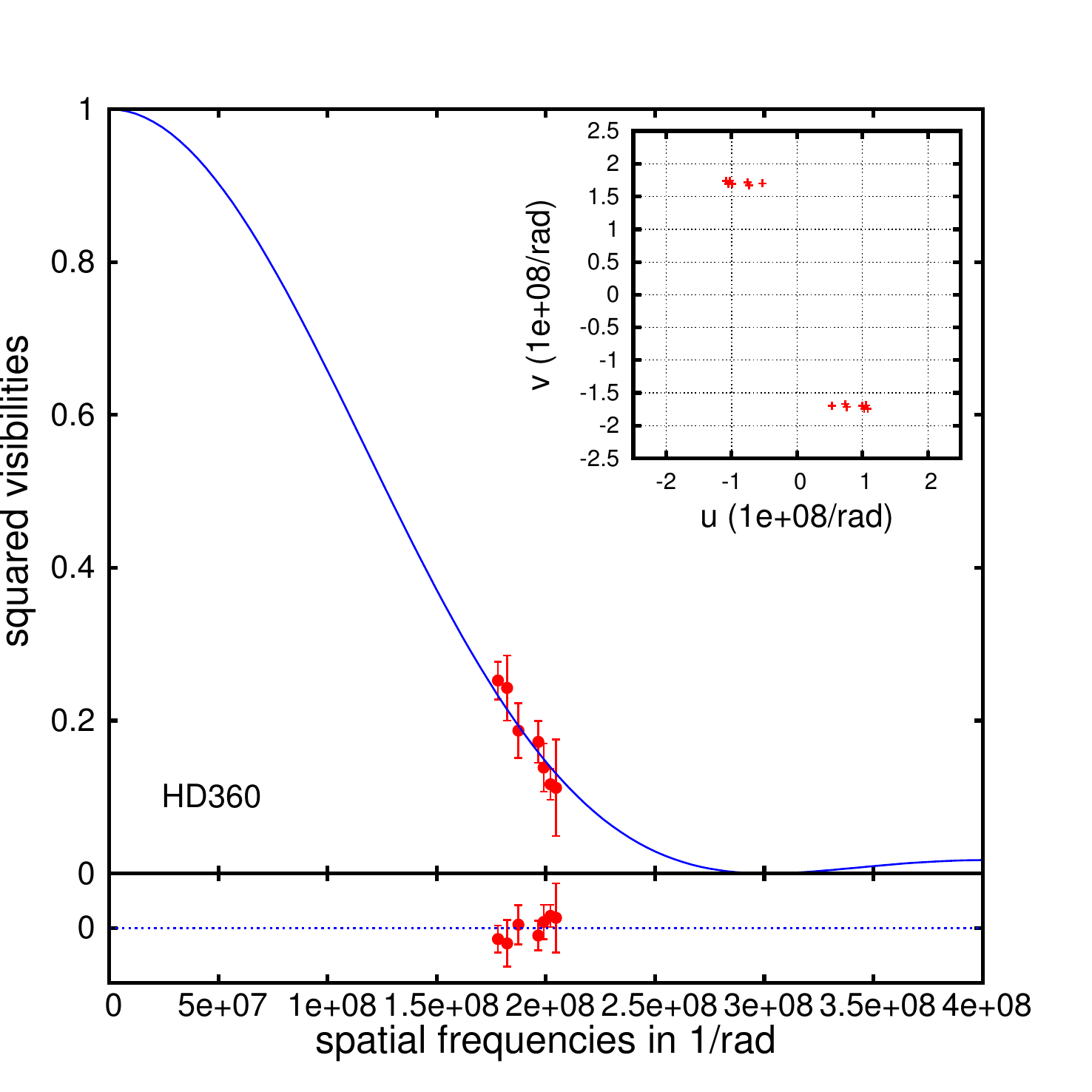}}
 \resizebox{0.35\hsize}{!}{\includegraphics[clip=true]{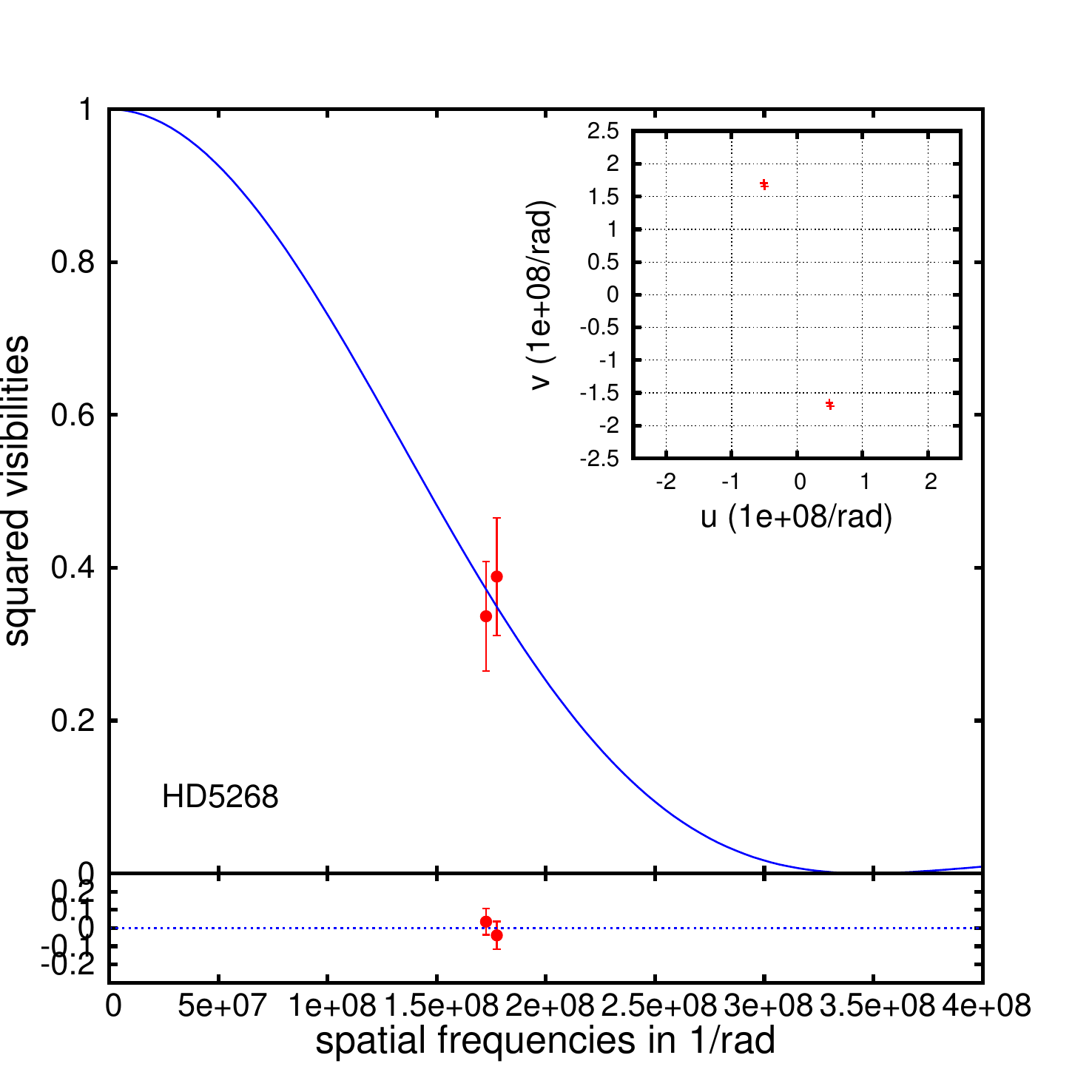}}
 \resizebox{0.35\hsize}{!}{\includegraphics[clip=true]{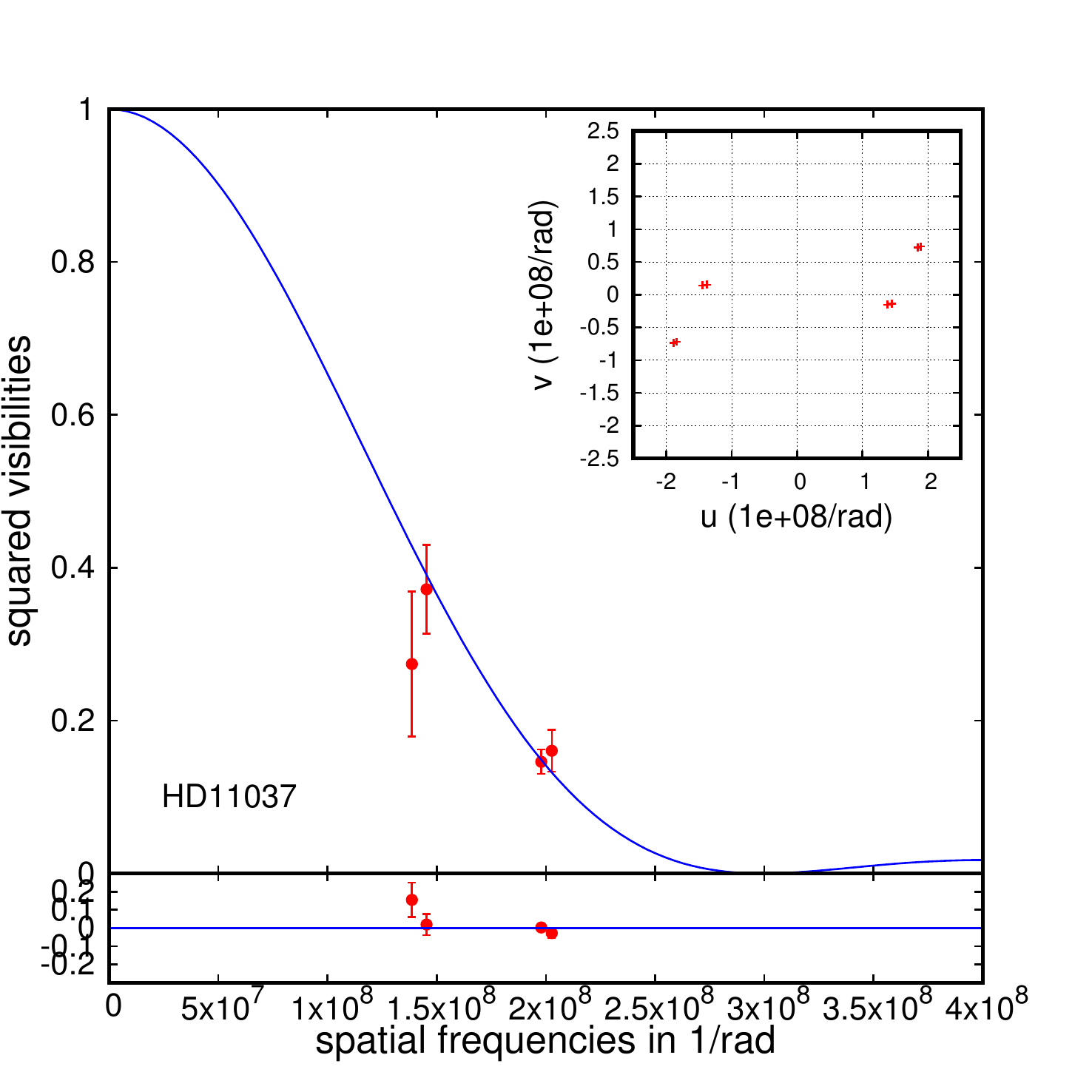}}
 \resizebox{0.35\hsize}{!}{\includegraphics[clip=true]{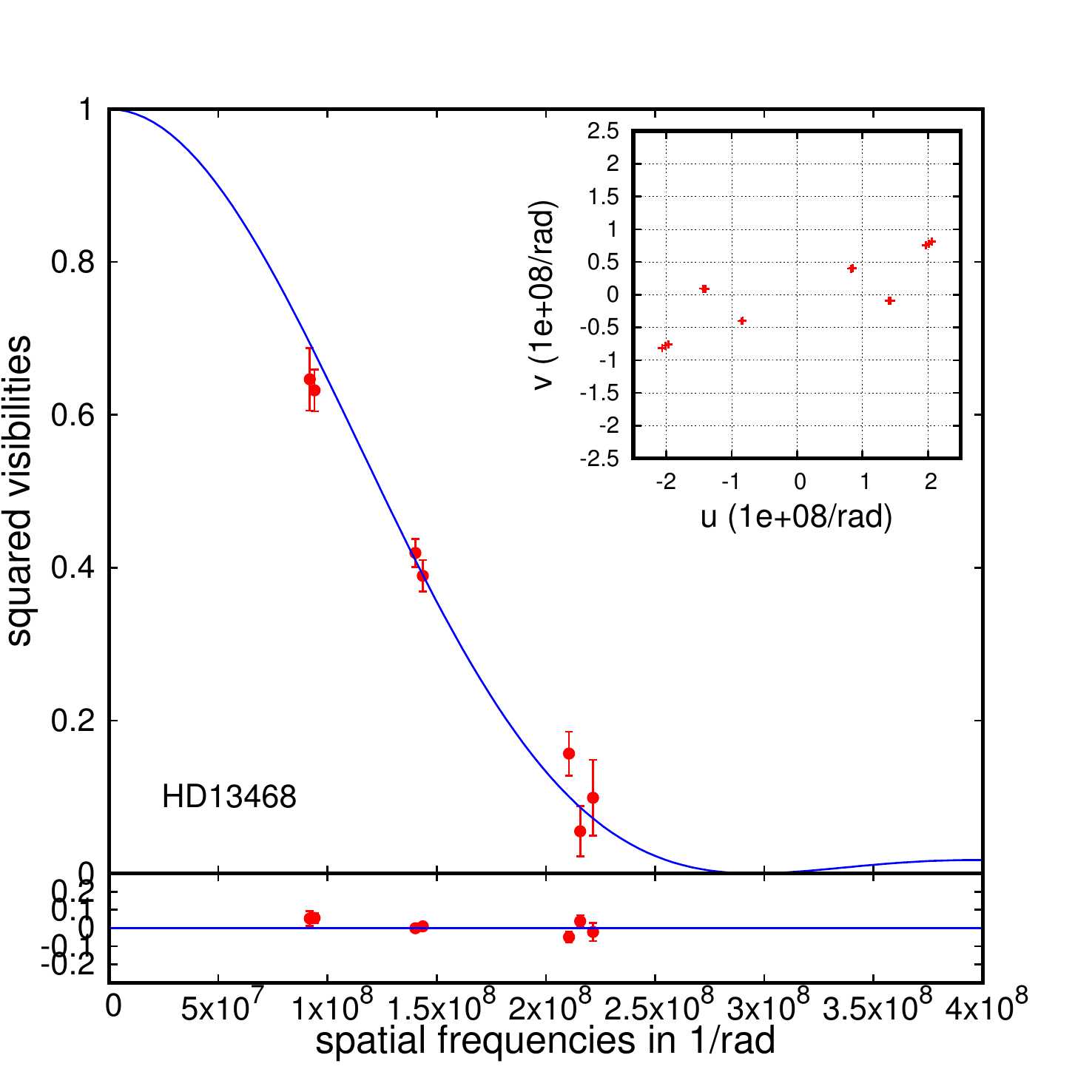}}
 \resizebox{0.35\hsize}{!}{\includegraphics[clip=true]{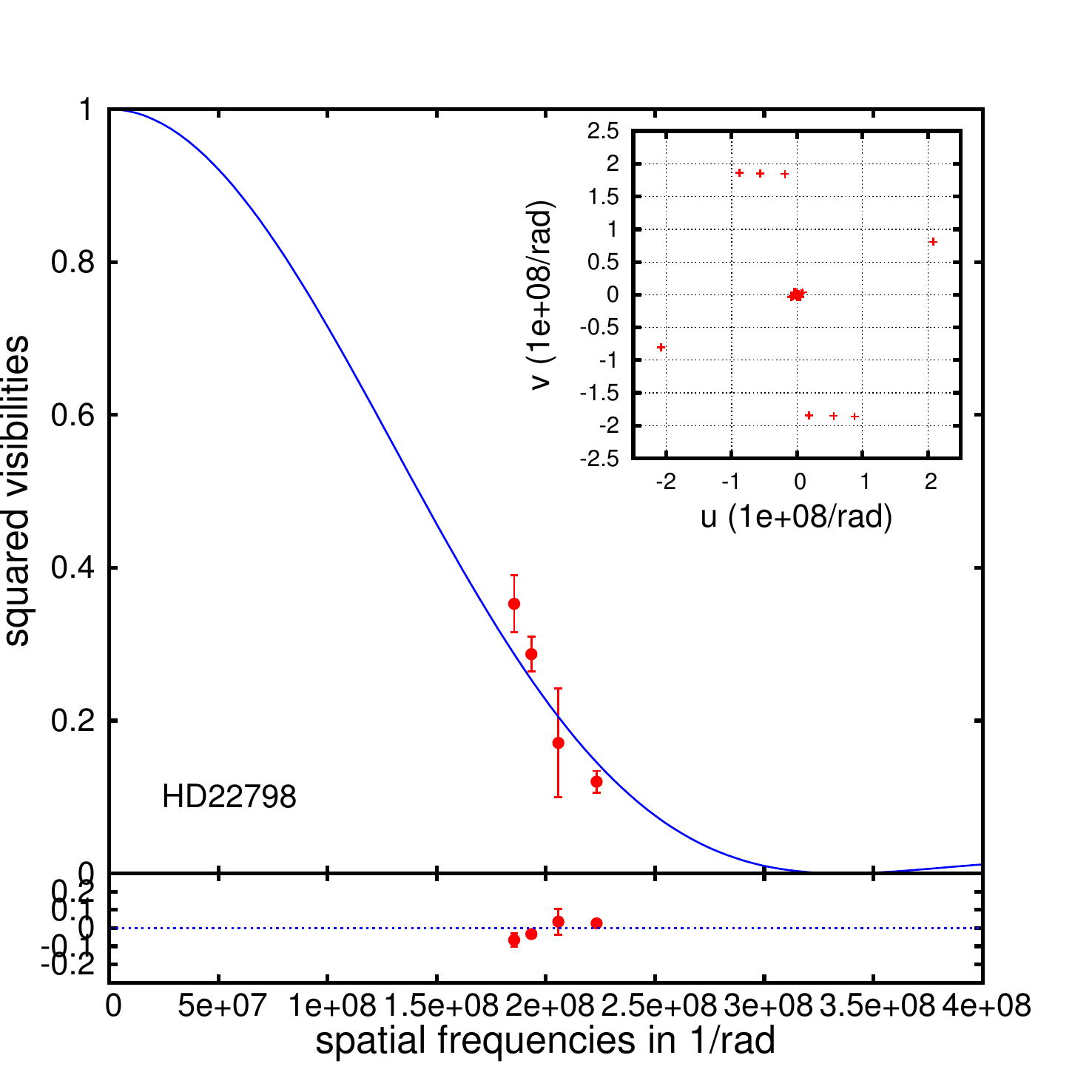}}
  \resizebox{0.35\hsize}{!}{\includegraphics[clip=true]{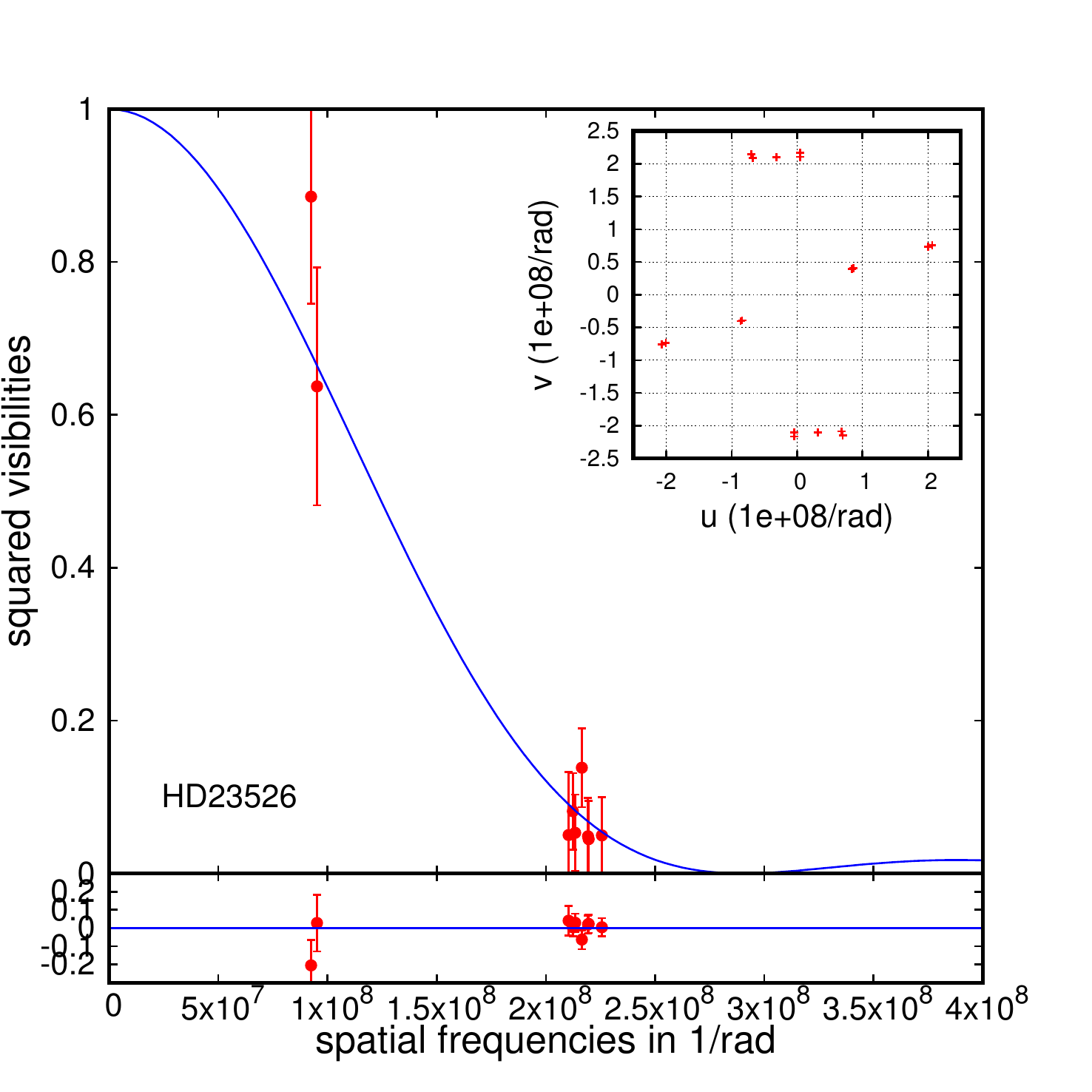}}
\caption {Squared visibility vs. spatial frequency for all stars in our sample with their corresponding statistical uncertainties (red dots). The solid blue lines indicate the best uniform-disk model obtained from the LITpro fitting software.} \label{fig.V2}
\end{center}
\end{figure*}

\begin{figure*}
 \begin{center}

    \resizebox{0.35\hsize}{!}{\includegraphics[clip=true]{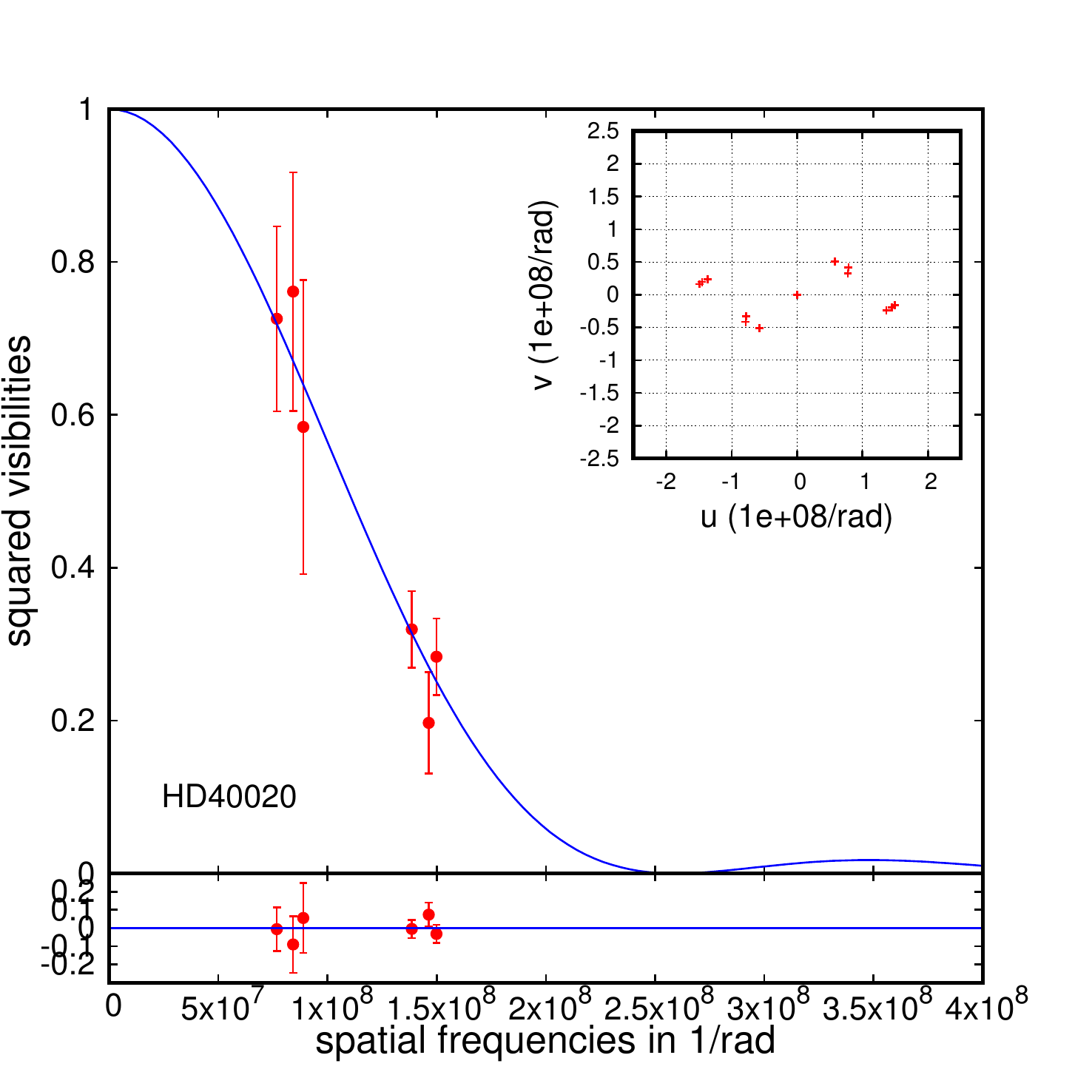}}
      \resizebox{0.35\hsize}{!}{\includegraphics[clip=true]{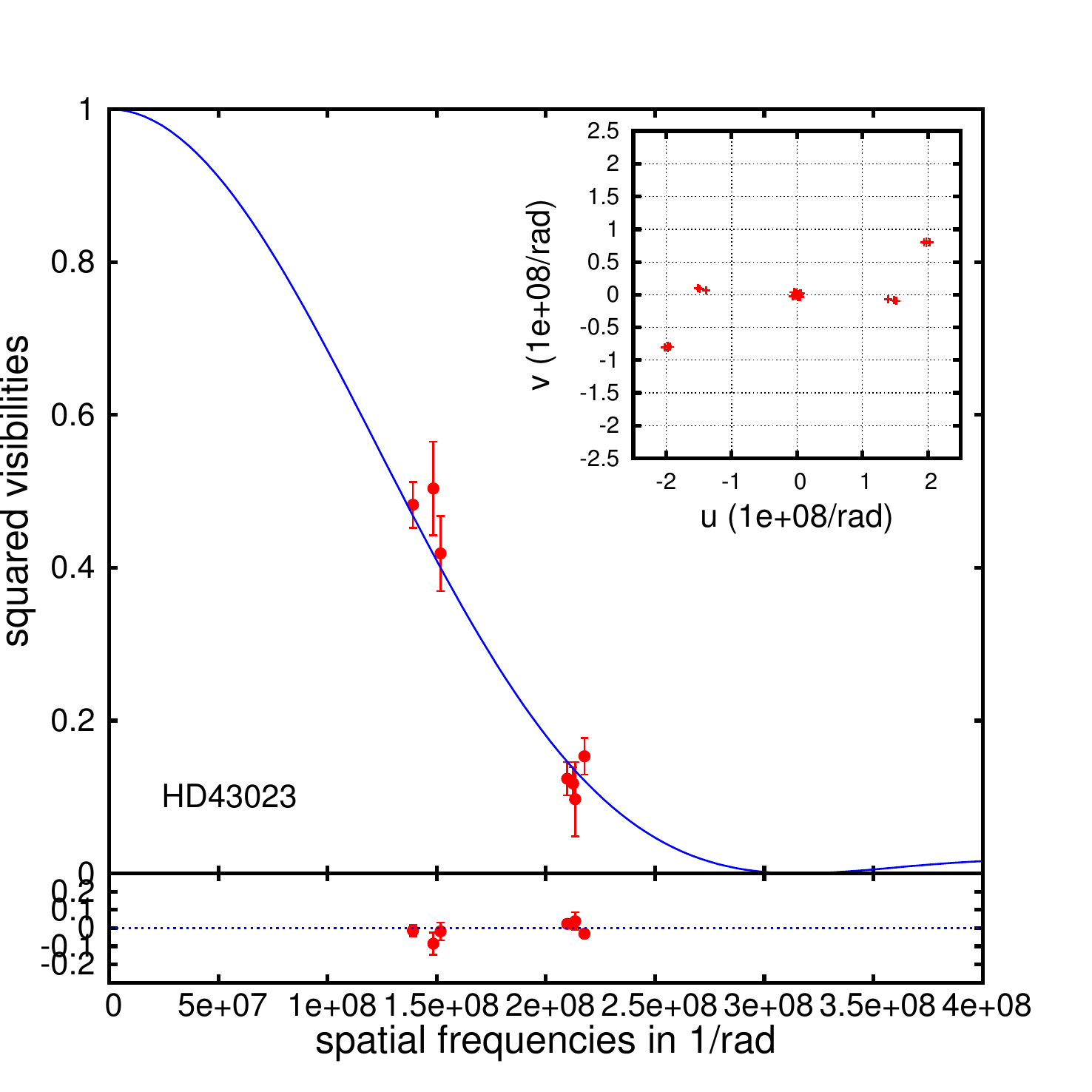}}
  
\caption {Squared visibility vs. spatial frequency for all stars in our sample with their corresponding statistical uncertainties (red dots). The solid blue lines indicate the best uniform-disk model obtained from the LITpro fitting software.} \label{fig.V2b}
\end{center}
\end{figure*}

\section{Table of the observing log}
\begin{table*}
\begin{center}
\caption[]{ Observing log. The columns list the date, the reduced Julian date ($RJD=JD-2400000$), the hour angle (HA), the minimum and maximum wavelengths over which the squared visibility is calculated, the projected baseline length Bp, and its orientation PA. The last column provides the calibrated squared visibility $V^{2}$ together with the statistic error on $V^{2}$ and the systematic error on $V^{2}$ (see text for details). Only the visibilities with a signal-to-noise ratio on the fringe peak higher than 1.7 were considered. The data are available on the Jean-Marie Mariotti Center \texttt{OiDB} service (Available at http://oidb.jmmc.fr).}
\label{Tabapx.1}
\begin{tabular}{lccccccccc}
\hline\hline
  &    Date             &   RJD   & HA  &  $\lambda_\mathrm{min}$ & $\lambda_\mathrm{max}$ & baseline & Bp &  PA    &$V^{2}_\mathrm{cal \pm stat \pm syst}$ \\
                       & [ yyyy.mm.dd ]&  [ days ] & [ h ] &             [ nm ]&                     [ nm ]  &  &   [ m ] & [ deg ] &                                                                  \\
                       
\hline
HD11037

&2013.10.27     &       56592.686       &       -2.60   &       705     &       725     &E2W2   &144.95 &       -111.33 &       0.160   $_\mathrm{\pm   0.027   \pm     0.003   }$\\
&2013.10.27     &       56592.686       &       -2.60   &       725     &       740     &E2W2   &144.95 &       -111.33 &       0.146   $_\mathrm{\pm   0.016   \pm     0.002   }$\\
&2014.10.19     &       56949.773       &       -1.03   &       695     &       715     &W1W2   &97.74  &       -83.73  &       0.274   $_\mathrm{\pm   0.095   \pm     0.002   }$\\
&2014.10.19     &       56949.793       &       -0.57   &       695     &       715     &W1W2   &102.47 &       -84.45  &       0.372   $_\mathrm{\pm   0.058   \pm     0.003   }$\\
\hline                                                                                                                                          
HD13468

&2014.08.27     &       56896.906       &       -1.66   &       690     &       710     &E1E2   &65.85  &       -115.41 &       0.632   $_\mathrm{\pm   0.027   \pm     0.003   }$\\
&2014.08.27     &       56896.906       &       -1.66   &       690     &       710     &       E2W2&155.13     &       -111.55 &       0.099   $_\mathrm{\pm   0.050   \pm     0.003   }$\\
&2014.08.27     &       56896.945       &       -0.65   &       690     &       710     &       E1E2 &64.31  &       -115.57 &       0.647   $_\mathrm{\pm   0.041   \pm     0.003   }$\\
&2013.11.02     &       56598.777       &       -0.51   &       705     &       725     &       W1W2&102.72     &       93.72   &       0.389   $_\mathrm{\pm   0.021   \pm     0.004   }$\\
&2013.11.02     &       56598.777       &       -0.51   &       705     &       725     &       W2E2&154.24     &       68.88   &       0.055   $_\mathrm{\pm   0.033   \pm     0.001   }$\\
&2013.11.02     &       56598.777       &       -0.51   &       725     &       740     &       W1W2&102.72     &       93.72   &       0.419   $_\mathrm{\pm   0.019   \pm     0.004   }$\\
&2013.11.02     &       56598.777       &       -0.51   &       725     &       740     &W2E2   &154.24 &       68.88   &       0.157   $_\mathrm{\pm   0.029   \pm     0.004   }$\\
\hline                                                                                                                                          
HD22798 
                                                                                                                                                                        
&2014.08.22     &       56892.002       &       -1.24   &       690     &       710     &       E2W2&156.25     &       -111.20 &       0.120   $_\mathrm{\pm   0.014   \pm     0.002   }$\\
&2014.10.20     &       56950.816       &       -1.83   &       695     &       715     &W2S2&130.75    &       -5.76   &       0.353   $_\mathrm{\pm   0.037   \pm     0.004   }$\\
&2014.10.20     &       56950.853       &       -0.91   &       695     &       715     &       W2S2&136.30     &       -16.83  &       0.287   $_\mathrm{\pm   0.023   \pm     0.004   }$\\
&2014.10.20     &       56950.890       &       -0.06   &       695     &       715     &       W2S2&145.01     &       -25.27  &       0.171   $_\mathrm{\pm   0.071   \pm     0.002   }$\\
\hline                                                                                                                                          
HD23526                                                                                                                                                                         
&2014.08.22     &       56891.974       &       -1.96   &       680     &       700     &       E2W2&151.53     &       -110.15 &       0.045   $_\mathrm{\pm0.050      \pm0.001        }$\\
&2014.08.22     &       56891.974       &       -1.96   &       700     &       720     &       E2W2&151.53     &       -110.15 &       0.053   $_\mathrm{\pm0.050      \pm0.001        }$\\
&2014.08.23     &       56892.985       &       -1.64   &       680     &       700     &       E1E2&65.66      &       64.98   &       0.637   $_\mathrm{\pm0.156      \pm0.002        }$\\
&2014.08.23     &       56892.985       &       -1.64   &       700     &       720     &       E1E2&65.66      &       64.98   &       0.886   $_\mathrm{\pm0.140      \pm0.002        }$\\
&2014.10.20     &       56950.797       &       -2.38   &       680     &       700     &       W2S2&149.38     &       -178.82 &       0.138   $_\mathrm{\pm0.052      \pm0.002        }$\\
&2014.10.20     &       56950.797       &       -2.38   &       700     &       720     &       W2S2&149.38     &       -178.82 &       0.050   $_\mathrm{\pm0.082      \pm0.001        }$\\
&2014.10.20     &       56950.834       &       -1.51   &       700     &       720     &       W2S2&150.78     &       171.40  &       0.081   $_\mathrm{\pm0.050      \pm0.001        }$\\
&2014.10.20     &       56950.872       &       0.59    &       680     &       700     &       W2S2&155.72     &       161.98  &       0.050   $_\mathrm{\pm0.050\pm0.001      }$\\
&2014.10.20     &       56950.872       &       0.59    &       700     &       720     &       W2S2&155.72     &       161.98  &       0.049   $_\mathrm{\pm0.050\pm0.001      }$\\
\hline                                                                                                                                          
HD360                                                                                                                                                                                   
&2013.08.26     &       56530.922       &       0.55    &       695     &       715     &       W2S2&144.26     &       -31.84  &       0.112   $_\mathrm{\pm   0.063   \pm     0.002   }$\\
&2013.08.26     &       56530.922       &       0.55    &       715     &       735     &       W2S2&144.26     &       -31.84  &       0.138   $_\mathrm{\pm   0.032   \pm     0.002   }$\\
&2013.07.24     &       56497.945       &       -1.00   &       690     &       710     &       W2S2&124.66     &       -17.34  &       0.252   $_\mathrm{\pm   0.025   \pm     0.006   }$\\
&2013.07.24     &       56497.971       &       -0.41   &       690     &       710     &       W2S2&131.20     &       -23.79  &       0.187   $_\mathrm{\pm   0.036   \pm     0.004   }$\\
&2013.07.24     &       56497.971       &       -0.41   &       710     &       730     &       W2S2&131.20     &       -23.79  &       0.243   $_\mathrm{\pm   0.043   \pm     0.005   }$\\
&2013.07.24     &       56498.003       &       0.35    &       690     &       710     &       W2S2&141.49     &       -30.45  &       0.117   $_\mathrm{\pm   0.020   \pm     0.003   }$\\
&2013.07.24     &       56498.003       &       0.35    &       710     &       730     &       W2S2&141.49     &       -30.45  &       0.172   $_\mathrm{\pm   0.027   \pm     0.005   }$\\
\hline                                                                                                                                          
HD40020                                                                                                                                                         
&2013.10.27     & 56592.860     &-2.63  &725 &  745     & E1E2&61.90    & 67.00   & 0.761 $_\mathrm{\pm0.156\pm   0.002}$\\
&2013.11.26     & 56622.863     &0.59   &725    & 745   & E1E2&65.35    & 62.21   &0.584  $_\mathrm{\pm0.192      \pm0.006}$\\
&2013.11.26     & 56622.955     &1.57   &705    & 725   & E1E2&54.87    & 48.71   & 0.726 $_\mathrm{\pm0.121      \pm0.008}$\\
&2014.10.22     & 56952.938     &-1.05  &700    & 720   & W1W2&98.39    & -80.11   &0.319 $_\mathrm{\pm0.050      \pm0.010}$\\
&2014.10.22     & 56952.962     &0.45   &700    & 720   & W1W2&103.90   & -82.36  &0.197  $_\mathrm{\pm0.067      \pm0.006}$\\
&2014.10.23     & 56953.976     & 0.04  &700    & 720   & W1W2&106.40   & -83.77   & 0.284        $_\mathrm{\pm0.050      \pm0.007}$\\

\hline                                                                                                                                          
HD43023                                                                                                                                                                                 
&2013.10.26     &       56591.903       &       -1.98   &       710     &       725     &       E2W2&153.65     &       -112.31 &       0.097   $_\mathrm{\pm   0.049   \pm     0.001   }$\\ 
&2013.10.26     &       56591.903       &       -1.98   &       725     &       740     &       E2W2&153.65     &       -112.31 &       0.124   $_\mathrm{\pm   0.022   \pm     0.001   }$\\
&2013.10.26     &       56591.919       &       -1.59   &       705     &       725     &       E2W2&155.62     &       -111.65 &       0.153   $_\mathrm{\pm   0.024   \pm     0.002   }$\\
&2013.10.26     &       56591.919       &       -1.59   &       725     &       740     &       E2W2&155.62     &       -111.65 &       0.118   $_\mathrm{\pm   0.021   \pm     0.001   }$\\
&2014.10.16     &       56947.000       &       -0.25   &       695     &       715     &       W1W2&104.65     &       -86.71  &       0.504   $_\mathrm{\pm   0.061   \pm     0.006   }$\\
&2014.10.16     &       56947.019       &       0.19    &       695     &       715     &       W1W2&106.99     &       -86.35  &       0.419   $_\mathrm{\pm   0.049   \pm     0.005   }$\\
&2014.10.18     &       56948.965       &       -0.96   &       695     &       715     &       W1W2&98.08      &       -87.20  &       0.482   $_\mathrm{\pm   0.030   \pm     0.004   }$\\
\hline                                                                                                                                          
HD5268                                                                                                                                                                                          
&2013.07.28     &       56501.961       &       -1.05   &       690     &       710     &W2S2   &124.28 &       163.39  &       0.388   $_\mathrm{\pm   0.077   \pm     0.009   }$\\
&2013.07.28     &       56501.961       &       -1.05   &       710     &       730     &       W2S2&124.28     &       163.39  &       0.336   $_\mathrm{\pm   0.072   \pm     0.008   }$\\
\hline
\end{tabular}
\end{center}
\end{table*}
\end{appendix}
\end{document}